\documentclass{aa}
\usepackage{graphicx}
\newcommand{\etal}{et al.}

\newcommand{\xmm}{{\it XMM-Newton}}
\newcommand{\rxte}{{\it RXTE}}

\begin{document}


\title{The long-term X--ray variability properties of AGN in the Lockman Hole
region}
\author{I. E. Papadakis\inst{1,2} \and  E. Chatzopoulos\inst{1} \and D.
Athanasiadis\inst{3} \and A. Markowitz\inst{4} \and I.
Georgantopoulos\inst{3}}
\offprints{I. Papadakis;  e-mail: jhep@physics.uoc.gr}
\institute{Physics Department, University of Crete, P.O. Box 2208, GR--710 03
Heraklion, Crete, Greece  \and IESL-Foundation for Research and Technology, 711
10 Heraklion, Crete,  Greece \and Institute of Astronomy \& Astrophysics,
National Observatory of Athens, I. Metaxa \& V. Pavlou, GR--152 36 P. Penteli,
Athens, Greece \and Center for Astrophysics and Space Sciences, University of
California, San Diego, M. C. 0424, LA Jolla, CA 92093-0424, USA}

\date{Received ?/ Accepted ?}

\abstract
{We present the results from a detailed X--ray  variability analysis of 66
AGN in the Lockman Hole, which have optical spectroscopic identifications.}
{We compare, quantitatively, their variability properties with the properties
of local AGN, and we study the ``variability -- luminosity" relation as a
function of redshift, and the ``variability -- redshift" relation in two
luminosity bins.}
{We use archival data from the last 10 \xmm\ observations of the Lockman Hole
field to extract light curves in the rest frame, 2--10 keV band. We use the 
``normalized excess variance" to quantify  the variability amplitude.  Using
the latest results regarding the AGN power spectral shape and its dependence on
black hole mass and accretion rate, we are able to compute model ``variability
-- luminosity" curves, which we compare with the relations we observe.}
{When we consider all the sources in our sample, we find that  their
variability amplitude decreases with increasing redshift and luminosity. These
global anti-correlations are less pronounced when we split the objects in
various luminosity and redshift bins. We do not find a significant correlation
between variability amplitude and spectral slope, $\Gamma$. The  ``variability
-- luminosity" relation that we detect has a larger amplitude when compared to
that of local AGN. We also find that, at a given luminosity, the variability
amplitude increases with redshift up to  z$\sim$1, and then stays roughly
constant.} 
{Our results imply that the AGN X--ray mechanism operates in the same way at
all redshifts. The accretion rate (in units of the Eddington limit) for the
objects in our sample increases from $\sim 0.25$, at z $\sim 0.5$, to 0.5  at z
$\sim 3$. It does not exceed unity even in the case of the most luminous AGN.
Their black hole mass  also increases with redshift. The upper limit we find is
consistent with the largest black hole masses found to date in the local
Universe. The black hole mass increase, and the decrease  of the rest frame
light curve's duration with increasing redshift, can explain the global
variability amplitude -- redshift/luminosity anti-correlations that we observe.
Among objects with the same luminosity, the black hole mass decreases and the
accretion rate increases with larger redshift. This effect explains the
increase of the variability amplitude up to redshift$\sim$1 (at fixed
luminosity bins). At higher redshifts, the decrease of the light curves length
affects the variability amplitude significantly, forcing it to remain
essentially constant.} 


\titlerunning{X-ray variability of AGN in the Lockman Hole}
\authorrunning{I. Papadakis~\etal}
\maketitle

\section{Introduction}

The X--ray variability properties of nearby AGN have been extensively studied
the last twenty years. One of the earliest, and best established so far, result
was that the variability amplitude decreases with increasing luminosity (e.g.
Barr \& Mushotzky 1986; Lawrence \& Papadakis 1993; Green \etal\ 1993; Nandra
\etal\ 1997; Turner \etal\ 1999; Markowitz \& Edelson 2004) although at
very low luminosities the anti-correlation breaks, and the variability amplitude
drops to nearly zero (e.g. Ptak et al. 1998). This ``variability -- luminosity"
anti-correlation may in fact be the result of a more fundamental ``variability
-- black hole (BH) mass" relation (e.g. Papadakis 2004;  O'Neil \etal\ 2005). If
that is the case, then one could  in principle get an  estimate of the BH mass
from a simple variability amplitude measurement (e.g. Nikolajuk, Papadakis \&
Czerny 2004; Nikolajuk \etal\ 2006).   Both the ``variability -- luminosity" and
the ``variability -- BH mass" anti-correlations are probably projections of a
``variability -- BH mass -- accretion rate" fundamental plane (M$^{\rm c}$Hardy
\etal\ 2006).

The ``variability BH mass estimation method" may be important in the case of
distant AGN, which have been detected in recent X--ray surveys and in many
cases they are too faint in other wavelengths for alternative BH mass
estimation methods to apply.  However, until recently, knowledge of the X-ray
variability properties of distant (z$>$0.1) AGN was scarce. The situation
changed the last few years, when data from deep X--ray surveys become
available, and techniques that are able to measure variability amplitudes for
low signal-to-noise sources were developed. 

Almaini \etal\ (2000) and Manners \etal\ (2002) presented the results from an
analysis of the X--ray variability of radio-quiet AGN selected from a deep {\it
ROSAT} survey and the {\it ROSAT} archive, respectively. They confirmed the
decline in variability amplitude with luminosity out to redshifts as large as
z$\sim$2. They also noted some evidence for an increase in the variability
amplitude at high redshifts. Paolillo \etal\ (2004) studied the X--ray
variability of sources detected in the Chandra Deep Field--South. They also
observed an anti-correlation with the luminosity of the sources and, in
agreement with the previous studies, they found that high-redshift objects have
larger variability amplitudes than expected from extrapolation of their low-z
counterparts. 

In the current study we measure the variability amplitude of AGN detected by
\xmm\ in the Lockman Hole (LH) field. \xmm\ has carried out its deepest survey
so far in the direction of this field. Mateos \etal\ (2005) studied the
spectral properties of  the 123 brightest  sources (i.e. sources more than 500
MOS+PN counts in the 0.2--12 keV band) in this field.  Recently, Mateos \etal\
(2007) presented the results from a detailed study of their flux variability
properties as well, using data from 16 of the currently available \xmm\
observations. They found that the variability amplitude  does not significantly
depend on redshift or X--ray luminosity. They also studied their spectral
variability properties, and found that spectral variability is less common that
flux variability and a lack of correlation between flux and spectral
variability.

In this work we study the variability properties of 66 objects with optical
spectroscopic identifications in the LH field, using data from the last 10
\xmm\ observations of the LH field only. These observations  were performed
over a  period of two months on  a  quasi-regular pattern. As a result, we were
able to construct  uniformly sampled, high signal-to-noise light curves for all
sources in our sample. Furthermore, different to all previous studies, we
extracted {\it rest frame} 2--10 keV band  light curves, and we compare,
quantitatively, our results with those from the local AGN sample of Markowitz
\& Edelson (2004). We also quantify the variability amplitude in a way
different to what has been used in most cases in the past (i.e. Almaini \etal\
2000, Manners \etal\ 2002, Mateos \etal\ 2007): we use the ``normalized excess
variance" a rather well known and frequently used estimator which is easy to
interpret. 

Our main aim is to study the correlation between the variability amplitude and
other source parameters like redshift, X--ray luminosity and X--ray spectral
slope and, in particular, to  investigate if and  how the
``variability--luminosity" and ``variability -- redshift" correlations changes
with redshift and luminosity, respectively.  Our results suggest that the
X--ray source in AGN operates in the same way at all redshifts, and allow us to
estimate how the average black hole mass and accretion rate of the AGN in our
sample change with redshift. 

\section{The \xmm\ observations and data reduction} 

\xmm\ observed the Lockman Hole field 17 times from April 2000 to December
2002. The total number of X--ray sources detected in this field is 268 (Mateos
\etal, 2005). Currently, 74 of these objects have been spectroscopically
identified as either Type I or Type II AGN. They constitute our initial sample
of sources. In order to investigate their variability properties we used data
from the last 10 \xmm\ observations only. A summary of these observations is
reported in Table~1. 

By choosing to work with observations which are spread over a period less than
two months we minimize the risk that potential changes in the response of the
\xmm\ detectors over large period of time will add to the observed   variations
in the light curves. Furthermore these 10 observations were done in a
quasi-regular temporal pattern. The first seven  were performed every ~2 days
over a period of ~14 days while the last three over a period of 10 days.
Consequently, the choice to use data from these observations only, minimizes
differences in the sampling pattern of the resulting light curves. 

\begin{table}
\begin{center}
\caption{Summary of the 10 LH  \xmm\ observations that we used in our study. In
the last column we list the ``good time interval" (GTI)  after the removal of
high background periods.} 
\begin{tabular}{ccc}
\hline
\hline
ObsID & Obs. Date & GTI (ksec) \\
\hline
0147510101 & 2002--10--15 & 91   \\
0147510801 & 2002--10--17 & 75   \\
0147510901 & 2002--10--19 & 90   \\
0147511001 & 2002--10--21 & 82   \\
0147511101 & 2002--10--23 & 50   \\
0147511201 & 2002--10--25 & 104  \\
0147511301 & 2002--10--27 & 65   \\
0147511601 & 2002--11--27 & 100  \\
0147511701 & 2002--12--04 & 99   \\
0147511801 & 2002--12--06 & 88   \\   
\hline                                    
\end{tabular}
\end{center}
\end{table}

\begin{table*}
\begin{center}
\caption{IDs, redshift, spectral and variability properties of the 66 AGN in the
Lockman Hole field that we study in this work. The first column lists the source
ID as given by Mateos et al. (2005) together with its RA and Dec.   Data listed
in columns 2, 9, and 10 are taken from Table 8 of Mateos \etal\ (2005). Numbers
in parenthesis in the third column correspond to the 3$\sigma$ upper limits on
$\sigma^2_{\rm NXS}$, in the case of the NV-sources. The values listed in column
9 correspond  to  the logarithm of the rest frame, 2--10 keV band luminosity of
each object. Column 8 indicates whether a light curve is considered as
``variable" (V) or ``non-variable" (NV) (for details see Section 3.1). Column 11
lists the \xmm\ detectors that were used to construct the respective light
curve. The full table is available in electronic form at the CDS.}
\begin{tabular}{ccccccccccc}
\hline
\hline
SourceID & z & $\sigma^2_{\rm NXS}$  &
$\log\sigma^2_{\rm NXS}$ & $\chi^2$/dof &
V/NV & log(L$_{\rm X}$) & $\Gamma$ & Instr. Used \\
\hline
5(10 52 43.3 +57 15 45.9) & 2.144 & 0.026$\pm$0.015 & -1.59$\pm$0.26 & 7.2/4  &  NV & 44.68 & 1.90 & PN \\
    &       & (0.045)  & & & & &  & \\
39(10 53 19.09 +57 18 53.6)  & 0.711 &-0.019$\pm$0.027 &  $-$  & 1.8/4  &  NV & 43.44 & 1.79 & M1+M2 \\
    &       & (0.081) & &  & & & & \\
41(10 51 19.14 +57 18 34.1) & 1.640 & 0.445$\pm$0.089 & -0.35$\pm$0.09 & 25.5/5 &  V  & 44.11 & 2.06 & PN \\
\hline                                    
\end{tabular}
\end{center}
\end{table*}

\subsection{Data reduction}

During the \xmm\  observations the EPIC PN and MOS cameras were operated in
full frame mode with a medium filter. The EPIC data were reprocessed with
{\small XMMSAS} version 6.5. We have selected PN photons with PATTERN$\leq 4$
(i.e. singles and doubles) and FLAG=0. In the case of the MOS cameras we have
selected events with  PATTERN $\leq$ 12 and FLAG=0.

Source counts were accumulated from a circular region of 18$^{\prime\prime}$
radius, centered around the coordinates of each source (as listed in Mateos
\etal\ 2005), in the energy range from 2 to 10 keV, in each source's  rest
frame. The background count rate, for all EPIC detectors, was  determined by
accumulating counts from six circular, source free regions, each one of radius
85$^{\prime\prime}$. Data from periods of high background were disregarded from
further analysis. Given the different redshifts, we constructed the appropriate
$2-10$ keV background light curves for each source individually. Finally, we
produced  PN, MOS1 and MOS2, background subtracted, rest frame $2-10$ keV light
curves, for each source in our sample. 

Since the \xmm\ EPIC cameras have different geometry, it was common that a
number of sources detected in each observation would fall near or inside CCD 
gaps, and/or close to the edge of at least one of the detectors. Furthermore,
in some cases we would  detect bad/hot pixels within a source's photon
extraction aperture in one (or more) of the EPIC cameras. For these reasons, we
inspected visually all sources in each PN, MOS1 and MOS2 image. We would also
check whether the background count rate we had estimated in an automatic way
would be representative of the local background around each source. 

As a result of this screening process, light curves from each EPIC detector
have a different number of points in most cases. Due to the geometry of the PN
detector, the number of sources falling on the PN CCD gaps is larger than the
MOS cameras. Consequently, the number of data points, N$_{\rm data}$, in the PN
light curves is usually smaller than the number of points in the MOS light
curves.

A reliable estimate of the variability amplitude requires the use of light
curves with: a) the largest possible N$_{\rm data}$ and b) the  highest
signal-to-noise ratio.  This ratio can be increased by combining counts from
all three EPIC cameras\footnote{One must be careful in combining data from the
three EPIC cameras, given the difference in their response. However, the
uncertainty introduced to our results by the difference in the average energy
of the (rest frame) 2--10 keV band photons in the EPIC PN, MOS1 and MOS2
cameras is smaller than that introduced by say the small number of points in
the light curves and the (unknown) stochastic nature of the underlying
variability process. For this reason, we use light curves with photons
accumulated from as many cameras as possible, as explained in the text.}. We
therefore added the background subtracted PN and MOS light curves to create
combined PN+MOS1+MOS2, PN+MOS1, PN+MOS2 and MOS1+MOS2 light curves.  Among the
individual and combined light curves of each source we would finally choose 
the one with the largest number of points. In the case there were more than
one, we would choose the light curve with data taken from as many as possible
EPIC detectors. 

The largest N$_{\rm data}$ value among all light curves is eight. Since N$_{\rm
data}$ (and hence the sampling pattern as well) should be roughly the same in
all of them, we required that N$_{\rm data}\geq 4$ (i.e. half of N$_{\rm
data,max}$). There are 66 (out of the 74) AGN whose light curves  satisfy this
criterion. These sources constitute our final sample. Table~2 lists their ID
number, redshift, X--ray spectrum slope, $\Gamma$, and rest frame, 2--10 keV
band luminosity, L$_{\rm X}$. In the last column of the same Table we list the
combination of the EPIC cameras that we used for each source.

There is just one object with N$_{\rm data}=4$ in our sample, and four with
N$_{\rm data}=8$. In most cases, N$_{\rm data}=$6. Due to this fact, the
sampling pattern of most  light curves is indeed similar, in the {\it
observer's frame}. However, when one takes into account the different 
redshifts of the sources in our sample, the light curves'  duration changes,
from one source to the other, in their {\it rest frame}. This redshift induced 
inhomogeneity in the temporal pattern of the light curves has to be taken into
account when we compare and interpret the variability properties of the low
and high-z objects in our sample.  

\section{Variability analysis method}

\subsection{Detection of intrinsic flux variations}

First we investigated which light curves show significant variations, using the
$\chi^{2}$ test. Since our sample consists of the brightest sources in the
catalogue of Mateos \etal\ (2005), there are more than 15 source photons in each
point of their light curves. This guarantees the applicability of Gaussian
statistics, and hence the reliability of the $\chi^{2}$ test results.  The results
($\chi^{2}/$degrees of freedom) are listed in the 7th column of Table 2. We accept
that a source shows significant flux variations if the probability of $\chi^{2}$
being larger than the obtained value, by chance, is smaller than 5\%. Column 8 on
the same table lists whether a source is variable (``V") or not (``NV"), according
to this criterion  \footnote {In the Appendix we discuss in detail the possibility
that the observed variations  may be affected significantly by instrumental
effects}. 

Sixteen objects, i.e. 24\% of the sources in the sample, turned out to be non
variable. In order to investigate this issue further, we  used the
Kolmogorov-Smirnov (K--S) test to compare the distribution functions of the 
redshift, N$_{\rm data}$ and mean count rate ($\bar{\rm CR}$) of the V and
NV-subsamples. 

We find that the redshift distribution functions of the two subsamples  are
almost identical. However, we get a different view when we consider the results
from the comparison of the $\bar{\rm CR}$ and N$_{\rm data}$ distributions. 
The NV-sources are systematically fainter than the V-sources. We find that 
$<\bar{\rm CR}_{\rm NV}>=7.1\times10^{-4}$ cnts/s and  $<\bar{\rm CR}_{\rm
V}>=2.2\times10^{-3}$ cnts/s. The K--S test suggests that the difference in
brightness is significant at the 96\% level. There are also less points in the
light curves of the NV-sources. We find that $<\rm{N}_{\rm data, NV}>=5.7$ as
opposed to $<\rm{N}_{\rm data, V}>=6.2$. The K--S test suggests that the
difference in the distribution of N$_{\rm data}$ is significant at the 99.94\%
level. We conclude that we do not detect intrinsic variations in the NV-sources
most probably because they are less bright and their light curves have fewer
points than the V-sources. 

Fig.~1 shows the rest frame, 2--10 keV light curve of the objects with ID
numbers V364 and V321. Their average count rate is similar, and in fact  there
are more points in the V364 light curve. Nevertheless, it is source  V321 which
shows significant variations  in its light curve. As Fig.~1 shows this is due
to just one point (which we have carefully examined to establish its
reliability). This result demonstrates clearly the importance of working with
light curves which have the largest possible number of points (and justifies
the way we constructed the light curves, as explained in the previous
section). 

\begin{figure}
\centering
\includegraphics[height=5cm,width=8.5cm]{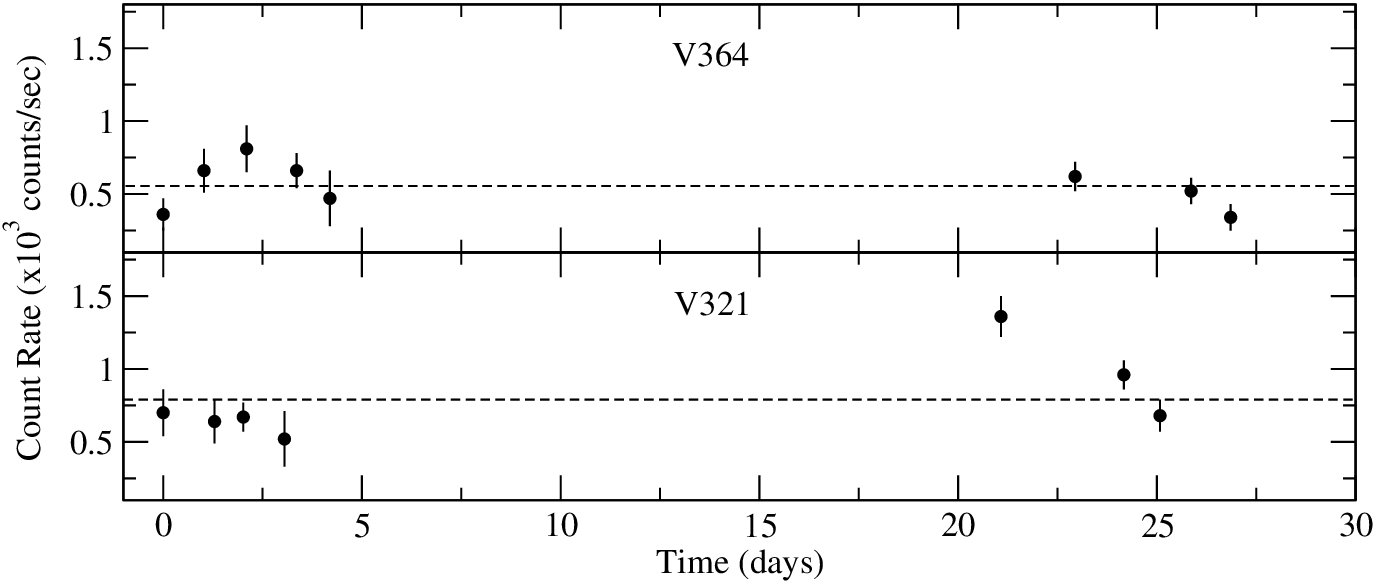}
\caption{The rest frame, 2--10 keV light curve of source  V364, which is a ``NV"
object, and V321, which belongs in the ``V" subsample. Time is measured in days
(in rest frame) since the first observation of each source.}
\end{figure}

\subsection{Determination of variability amplitude}

We use the  normalized excess variance, $\sigma^{2}_{\rm NXS}$ (e.g. Nandra
\etal\ 1997) as a measure of the intrinsic variability amplitude of the light
curves. This estimator is defined as,

\begin{equation}  
\sigma^{2}_{\rm NXS}=\frac{S^{2}-<\sigma_{err}^{2}>}{<x>^2},  
\end{equation}

\noindent where $<x>$, and $S^{2}=(1/\rm {N_{data}})\sum_{i=1}^{\rm
N_{data}}(x_{i}-<x>)^2$ are the mean and variance of the light curve, while 
$<\sigma^2_{\rm{err}}>=(1/{\rm N_{data}})\sum_{i=1}^{{\rm
N_{data}}}\sigma^{2}_{err,i}$ is the average contribution of the Poisson noise
process to the observed scatter around the mean. Its square root ($F_{\rm
var}$) indicates the average variability amplitude of a source as a fraction of
its light curve mean.

The spread of the observed $\sigma^2_{\rm NXS}$ values around the intrinsic,
``true" $\sigma^2$ value (i.e. the ``error" of $\sigma^2_{\rm NXS}$) should
decrease with increasing N$_{\rm data}$. This spread also depends on the 
stochastic nature of the process underlying the AGN X--ray variability (see
Vaughan \etal\ 2003 for a detailed discussion on this issue). In fact, even  if
a source is not intrinsically variable, the  $\sigma^2_{\rm NXS}$ estimates
based on  light curves from observations performed at different times will not
be identical, due to the presence of the Poisson noise alone. 

It is not easy to estimate a priori how does the error on  $\sigma^2_{\rm NXS}$
depend on N$_{\rm data}$ and on the stochastic nature of the variability
mechanism. On the other hand, the error due to Poisson noise can be estimated
using the following equation (Vaughan \etal\ 2003),

\begin{equation}
{\rm err}(\sigma^{2}_{\rm NXS})=\sqrt{\frac{2} {\rm N_{data}} ( \frac
{<\sigma^{2}_{\rm err}>} {<x>^2} )^2+ \frac {<\sigma^{2}_{\rm err}>} {\rm
N_{data}} \frac{4\sigma^{2}_{\rm NXS}}{<x>^2}}.
\end{equation}

\noindent This is useful in order to  estimate upper limits on the intrinsic
excess variance in the case when we do not detect significant variations. 

Columns 3 and 4 in Table~2 list the excess variance, and its error, for all the
sources in our sample. In the case of the NV-sources, $\sigma^2_{\rm NXS}$ is
consistent with zero, as expected. For these sources, the numbers in
parenthesis in the third column correspond to the $3\sigma$ upper limits, i.e. 
$\sigma^2_{\rm NXS}+3{\rm err}(\sigma^{2}_{\rm NXS})$. In columns 5 and 6 we
list the logarithm of $\sigma^2_{\rm NXS}$ and its error, estimated using the
usual ``propagation of errors" formula. 

\section{Correlation of variability amplitude with source parameters}

We now proceed to study the correlation between the excess variance and  source
parameters such as the redshift, L$_{\rm X}$, and $\Gamma$. In order to
quantify the correlation between two parameters, we use the  Kendall's  $\tau$
rank correlation coefficient. We accept a correlation to be significant if the
null hypothesis probability, P$_{\rm null}$, is less than 5\%. 

\begin{figure}
\centering
\includegraphics[height=10.5cm,width=8.5cm]{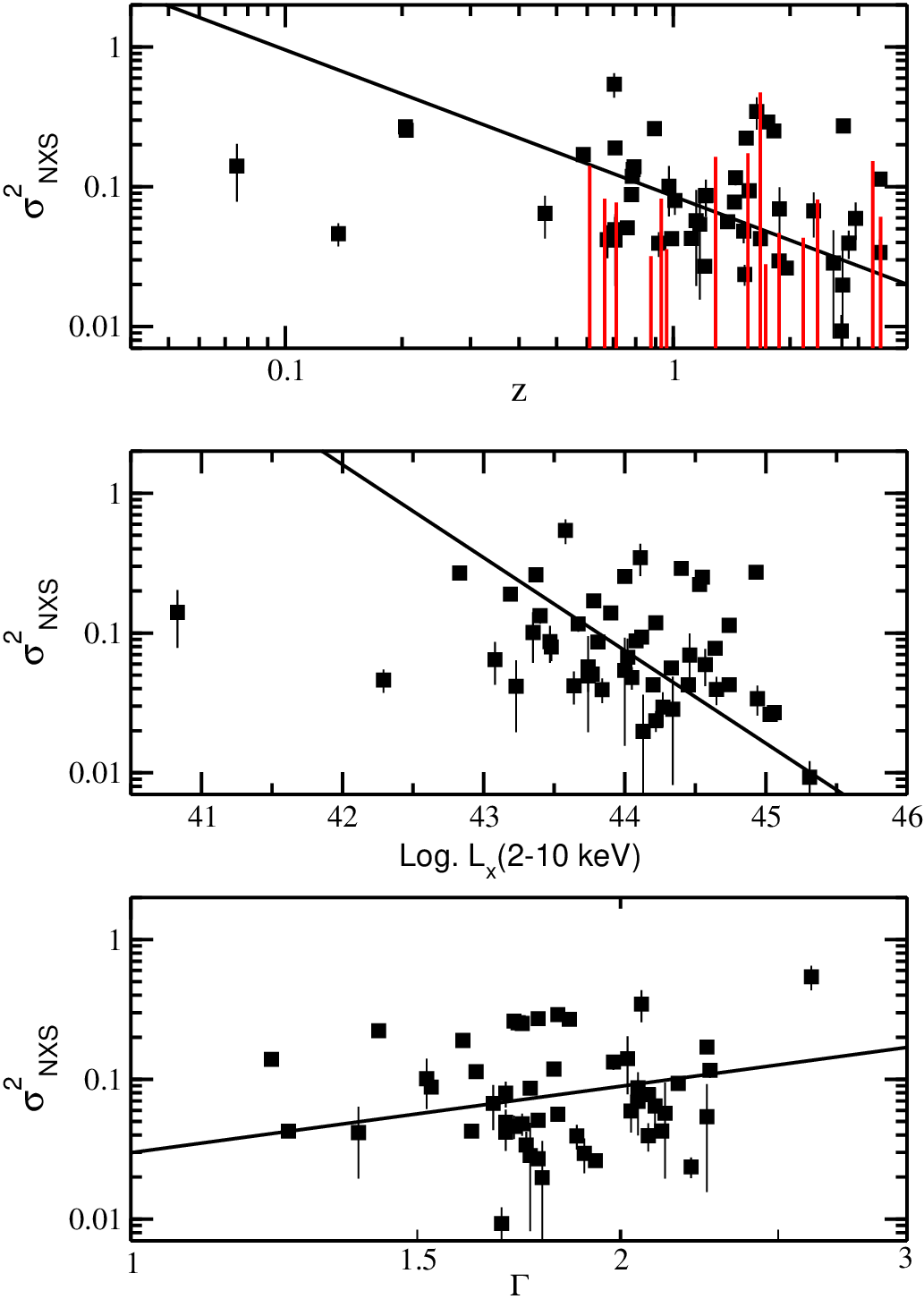} \caption{Plot of 
$\sigma^2_{\rm NXS}$ as a function of redshift (top panel), logarithm of the
rest-frame 2--10 keV luminosity (middle panel) and  spectral slope value
(bottom panel), for the 50 variable sources in our sample. The vertical gray
lines in the top panel show the $3\sigma$ upper limits on $\sigma^2_{\rm NXS}$
in the case of the NV-sources.}
\end{figure}

Filled squares in the upper panel of Fig.~2 show the ``$\sigma^2_{\rm NXS}$ vs
z" plot for the 50 variable sources in our sample. Errors are estimated using
equation (2). A trend of decreasing variability amplitude with increasing
redshift can be seen. When we consider the [log($\sigma^2_{\rm NXS}$),
log(z)] data for the variable sources only,  we find that $\tau=-0.2$, a
result which confirms the significance of the ``variability -- z"
anti-correlation (P$_{\rm null}=0.04$). 

The vertical grey lines in the same panel of Fig.~2 show the 3$\sigma$  upper
limits on  $\sigma^2_{\rm NXS}$ in the case of the NV-sources. They are so
large that they do not impose any significant constrains on the
``$\sigma^2_{\rm NXS}$ -- z" relation. We need better light curves for these
objects (in terms of signal-to-noise ratio and N$_{\rm data}$) to  constrain
this relation better. The same holds true for all relations we study below. For
that reason, we do not consider hereafter the NV-sources in our analysis. 

The solid line in the same figure shows the best fitting,  ``ordinary least
squares bisector" line to the data in log-log space (the best fitting 
parameter values were estimated as in Isobe \etal\ 1990).  We find that a
relation of the form $\sigma^2_{\rm NXS}\propto z^{-\alpha}$, with
$\alpha=-1.05\pm 0.10$ fits the data well. 

Next, we examined the relation between  variability amplitude and L$_{\rm X}$.
The middle panel in Fig.~2 shows the ``$\sigma^2_{\rm NXS}$ vs log(L$_{\rm X}$)"
plot for the V-objects in our sample. Given the correlation between L$_{\rm X}$
and redshift, it is not surprising that we an anti-correlation between
variability amplitude and luminosity can be seen. Kendall's $\tau$ 
suggests that the anti-correlation between log($\sigma^2_{\rm NXS}$) and
log(L$_{\rm X}$) is significant ($\tau=-0.22$, P$_{\rm null}=0.023$). The
solid line shows the best fitting $\sigma^2_{\rm NXS}\propto {\rm
L_X}^{-\alpha}$ line ($\alpha=-0.66\pm 0.12$), which fits the data rather well.

In the bottom plot we show  the ``$\sigma^2_{\rm NXS}$ vs $\Gamma$" relation
for the V-sources in our sample. One could perhaps suggest a positive
correlation between the two quantities, but Kendaull's $\tau$ in this case does
not support this claim ($\tau=0.03$, P$_{\rm null}=0.74$). The best fitting 
$\sigma^2_{\rm NXS}\propto \Gamma^{-\alpha}$ model to the data ($\alpha=1.58\pm
1.14$) is also shown with the solid line. The uncertainty  is so large that the
possibility of $\alpha=0$ (i.e. of no correlation between $\sigma^2_{\rm NXS}$
and $\Gamma$) cannot be excluded. 

The ``$\sigma^2_{\rm NXS}$ -- log(L$_{\rm X}$)" anti-correlation that we find is
in agreement with what has been observed in nearby AGN, both on long time scales
(e.g. Papadakis 2004; Markowitz \& Edelson, 2004) and short time scales (e.g. O'
Neil \etal\ 2005, and references therein). In fact our results  indicate that
the decline in variability amplitude with luminosity holds out to redshifts as
high as $\sim 3$. This  decline can also explain the ``$\sigma^2_{\rm NXS}$ --
z"  anti-correlation: as the more distant objects are systematically more
luminous, it is not surprising that they are also less variable. However, it is
important to note here that, although the correlations we detect are
statistically significant, they are also driven by a few individual points. For
example, if we omit the high-z source with $\sigma^2_{\rm NXS}<0.01$ then
P$_{\rm null}>0.05$ in the case of the [log($\sigma^2_{\rm NXS}$), log(z)]
anti-correlation. We reach the same conclusion if we omit the two pints with the
smallest $\sigma^2_{\rm NXS}$ in the case of the [log($\sigma^2_{\rm NXS}$),
log(L$_{\rm X}$)] anti-correlation. Consequently, one has to be careful
regarding the reality of these correlations.

\section{Comparison with nearby AGN}

In this section we present the results from a quantitative comparison between
the ``variability -- luminosity" relation as determined for nearby AGN and the
V-objects in our sample, grouped in various redshift bins. We use the
Student's-$t$ test to compare the mean of two sample distributions. We consider
two means as being significantly different if P$_{\rm null}$ is smaller than
5\%.

In order to determine the ``log($\sigma^2_{\rm NXS}$) vs log(L$_{\rm X}$)" 
relation for the AGN in the local universe we considered the 12 objects in the
sample of Markowitz \& Edelson (2004). Object names and the dates of the \rxte\
observations that were used to extract light curves are listed in Table 1 of
Markowitz \& Edelson (2004). Details of the data reduction can also be found in
the same paper.  

Filled squares in all panels of Fig.~3 show the [log($\sigma^2_{\rm NXS}$), 
log(L$_{\rm X}$)] data of these AGN. Excess variances were estimated as
described in Section 3.2. In the case when there were more than one light
curves for an object, we plot the straight mean of the individual
$\sigma^2_{\rm NXS}$ values.  We used 1.6 days binned, 2--12 keV light curves,
which were 23.2 days long. This is equal to the average, rest frame light curve
length of the V-sources in our sample. As for L$_{\rm X}$, we adopted the
$2-12$ keV luminosity measurements of Markowitz \& Edelson (2004). This is
slightly different to the energy band we chose in this work, but this should 
not affect our results significantly.  

The open circles in the top, middle and bottom panel of Fig.~3 show the 
[log($\sigma^2_{\rm NXS}), \log(\rm{L}_{\rm X})$] data of the V-sources  with
0.2$<$z$<$1 (the ``Low"-redshift bin; 18 objects), 1$<$z$<$2 (``High"-redshift
bin; 21 objects) and 2.3$<$z$<$3.4 (``very High", or vHigh, redshift bin; 9
objects). The average redshift of the objects in the  Low-z, High-z and v-High
bins is $<\rm{z_{Low-z}}>=0.7$, $<\rm{z_{High-z}}>=1.48$ and
$<\rm{z_{vHigh-z}}>=2.85$.

In order to reduce the scatter of the Low-z data points in Fig.~3, we sorted
them in order of increasing luminosity,  we considered three bins of 6 points
each, and we estimated their average log($\sigma^2_{\rm NXS}$) and log(L$_{\rm
X})$ values (shown with filled circles in the top panel of Fig.~3). Similarly, 
filled circles in the middle panel show the mean log($\sigma^2_{\rm NXS}$) and
log(L$_{\rm X})$ values of the data in three bins (each of 7 points) into which
we grouped the High-z objects (in order of increasing luminosity). We also
binned the vHigh-z objects in two groups and we estimated their average
log($\sigma^2_{\rm NXS}$) and log(L$_{\rm X}$) values. The results are shown
with the filled circles in the bottom panel of Fig.~3. 

On average, Low-z and nearby AGN have similar X--ray luminosity. We find that 
$<\log(\rm{L}_{\rm X,Low-z})>=43.62$ as opposed to $<\log(\rm{L}_{\rm X,near
by})>=43.31$ (P$_{\rm null}=0.21$). On the other hand, the High-z and vHigh-z
AGN are significantly more luminous than the nearby AGN: $<\log(\rm{L}_{\rm
X,High-z})>=44.24$ and $<\log(\rm{L}_{\rm X,vHigh-z})>=44.63$  (P$_{\rm
null}=4\times 10^{-4}$, and P$_{\rm null}=7.3\times 10^{-4}$, respectively). 
If both the LH and the nearby AGN  followed the same ``$\sigma^2_{\rm NXS}$ --
$\log(\rm{L}_{\rm X})$"  relation, we would expect the Low-z objects to be as
variable as the nearby AGN, and the  High-z and vHigh-z AGN to be significantly
less variable. However, this is not the case. 

The average variability amplitude of the nearby and the Low-z AGN are 
$<\log(\sigma^2_{\rm NXS,nearby})>=-1.45$ and  $<\log(\sigma^2_{\rm
NXS,Low-z})>=-0.97$. We find that their difference is significant (P$_{\rm
null}=0.003$). Even the High-z sources are significantly more variable than the
nearby AGN ($<\log(\sigma^2_{\rm NXS,High-z})>=-1.13$, P$_{\rm null}=0.031$),
while the vHigh-z objects are at least as variable as the nearby AGN, despite
the fact that they are much more luminous ($<\log(\sigma^2_{\rm
NXS,vHigh-z})>=-1.34$, P$_{\rm null}=0.57$). 

We conclude that, at a given X--ray luminosity, the z$>$0.2, V-sources in our
sample are systematically more variable than the nearby AGN. This is in
agreement with the results of Almaini \etal\ (2002), who found that the z$>0.5$
AGN do not show the anti-correlation with luminosity seen in local AGN, and the
results of Paolillo \etal\ (2004), who found that  high-z objects have larger
variability amplitudes than expected from extrapolations of their low-z
counterparts. 

\begin{figure}
\centering
\includegraphics[height=10.5cm,width=8.5cm]{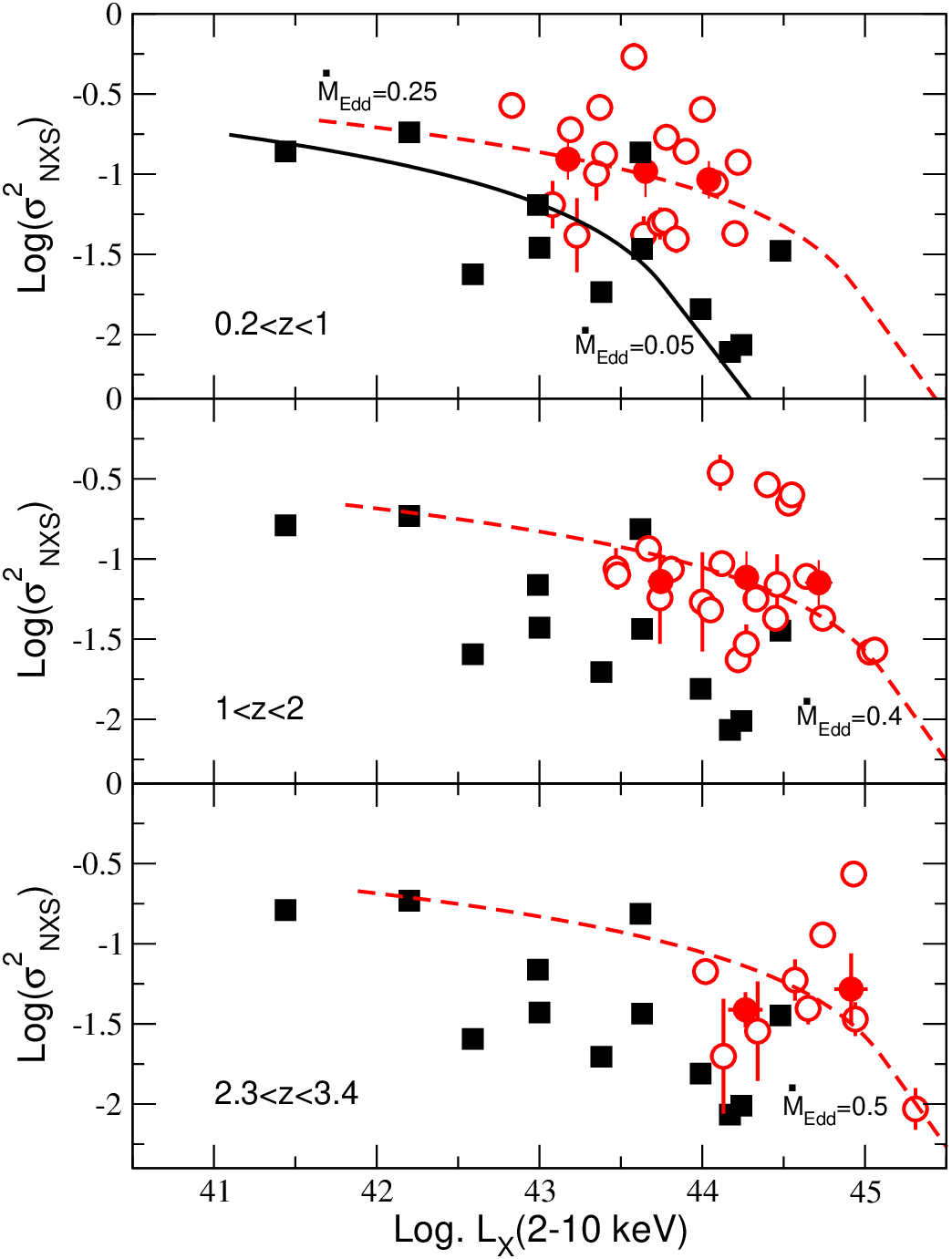}
\caption{The ``variability -- luminosity" plot for the nearby AGN (filled
squares in all plots), the Low-z, High-z and vHigh-z objects in the V-subsample
(open circles in the top, middle and bottoms panels, respectively). Filled,
grey circles correspond to the mean variability amplitude and luminosity, when
data are grouped in various bins as explained in the text. The solid line in
the top panel and the long-dashed lines in all panels show the model
``variability amplitude --  log(L$_{\rm X}$)" lines, estimated as explained in
Section 6.}
\end{figure}

\section{Impact of BH mass and accretion rate to the variability -- luminosity 
relation}

Let us assume that the X--ray variability mechanism is the same in all AGN,
irrespective of their redshift. This assumption implies that the power spectral
density, PSD, has the same shape in all of them. Recent work, based on 
detailed analysis of high quality \rxte\ and \xmm\ light curves, has shown that
the AGN PSDs have a power-law shape of slope $\sim -2$ for the 2--10 keV band
(as opposed to $\sim -2.7$ for softer bands), which changes to a slope of $\sim
-1$ below a so-called  ``break frequency", $\nu_{bf}$ (e.g. Markowitz \etal,
2003; M$^{\rm c}$Hardy \etal, 2004). 

The excess variance estimated from a light curve of length $T$, is an
approximate estimator of the following integral, 

\begin{equation}  
\sigma^2_{\rm NXS}=\int_{\nu_{lf}}^{\infty} PSD(\nu)d\nu ,
\end{equation}

\noindent where $\nu_{lf}=1/T$. If, as we assumed above,
$PSD(\nu)=A(\nu/\nu_{bf})^{-1}$ ($\nu<\nu_{bf}$) and 
$PSD(\nu)=A(\nu/\nu_{bf})^{-2}$ (when $\nu>\nu_{bf}$), then,  

\begin{equation}  
\sigma^2_{\rm NXS}=PSD_{amp}[\ln (\nu_{bf})- \ln (\nu_{lf})+1],  
\end{equation}

\noindent where $PSD_{amp}=A\nu_{bf}\sim 0.02$ (Papadakis 2004). In the case
when  $\nu_{lf}>\nu_{bf}$, 

\begin{equation}   
\sigma^2_{\rm NXS}=(PSD)_{amp} (\nu_{bf}/\nu_{lf}). 
\end{equation}

\noindent Recently, M$^{\rm c}$Hardy \etal\ (2006)  have demonstrated that
$\nu_{bf}$ depends on both the BH mass, M$_{\rm BH}$, and the accretion rate,
$\dot{\rm m}_{\rm Edd}$ (measured in Eddington units), as follows,

\begin{equation}   
\nu_{bf}=0.029\eta \dot{\rm m}_{\rm Edd}({\rm M_{BH}}/10^6\rm{M_{\odot}})^{-1}, 
\end{equation}

\noindent where $\eta$ is the efficiency of the mass to energy conversion
(hereafter we assume $\eta=0.1$). One can now use the above equation to
substitute $\nu_{bf}$ in equations 4 and 5 in order to relate $\sigma^2_{\rm
NXS}$ with M$_{\rm BH}$ and $\dot{\rm m}_{\rm Edd}$. The resulting relation
implies that for a given M$_{\rm BH}$, $\sigma^2_{\rm NXS}$ increases with
increasing accretion rate, while, among all AGN with the same $\dot{\rm m}_{\rm
Edd}$, objects with smaller M$_{\rm BH}$ should have larger variability
amplitude. 

The $\sigma^2_{\rm NXS}=f(M_{\rm BH},\dot{\rm m}_{\rm Edd})$ relation can be
transformed to a $\sigma^2_{\rm NXS}=f(\rm{L}_{\rm X}$) relation if we use the
fact that the bolometric luminosity, L$_{\rm bol}$, is given by L$_{\rm
bol}=1.3\eta\dot{\rm m}_{\rm Edd}10^{39}(\rm{M}_{\rm bh}/\rm{M}_{\odot}$) 
erg/s, and we adopt an X--ray luminosity to  L$_{\rm bol}$ conversion factor. 
To this end, we adopted the Marconi \etal\ (2004) prescription,

\begin{equation}   
\rm {log(L_X)}=\log(L^\ast)-1.54-0.24\ell-0.012\ell^{2}-0.0015\ell^{3},
\end{equation}

\noindent where $\ell=\rm logL^\ast-12$, L$^\ast$ being the bolometric
luminosity in solar luminosity units. 

To summarize, it is possible to derive a model ``$\sigma^2_{\rm NXS}$ vs
L$_{\rm X}$" relation, if we assume that: 1) all AGN, at all redshifts, vary
like the local AGN, 2) the efficiency is the same in all objects, and 3) a
bolometric to X--ray luminosity conversion relation. In addition, the model
takes into account the different {\it rest frame} length of the derived light
curves (``the redshift induced inhomogeneity" we mentioned in Section 2).
Although $T_{\rm obs}$ is similar in all light curves, $\nu_{lf}$ in equations
(4) and (5) should be equal to  $1/T_{\rm rest frame}$, where $T_{\rm rest
frame}=T_{\rm obs}/(1+z)$. 

The solid line in the upper panel of Fig.~3 shows such a model 
``log($\sigma^2_{\rm NXS}$) vs log(L$_{\rm X}$)" curve in the case when
$\dot{\rm m}_{\rm Edd}=0.05$ and  $\nu_{lf,{\rm nearby}}=1/23.2$ days$^{-1}$.
Clearly, this line describes rather well the ``variability -- luminosity"
relation for the nearby AGN. This is not surprising given the fact that the
model PSD shape that we have assumed, and the ``$\nu_{bf}$ vs BH mass and
accretion rate" relation that we use, have resulted from the study of objects
which are all included in our nearby AGN sample. 

The long dashed line in the same panel show the model ``log($\sigma^2_{\rm
NXS}$) vs log(L$_{\rm X}$)" relation in the case  when $\dot{\rm m}_{\rm
Edd}=0.25$ and  $\nu_{lf}=1/29.6$ days$^{-1}$ (29.6 days is the average, rest
frame duration of the Low-z light curves). This is entirely consistent with the
mean [log($\sigma^2_{\rm NXS}), \rm{log(L}_{\rm X})$] data of the Low-z
objects. 

The model curves plotted in the upper panel of Fig.~3 can be used to explain
the differences we observe between the nearby and Low-z AGN.  For example, let
us consider the objects with log(L$_{\rm X})=44$. According to the model, in
the local Universe these are objects with log(M$_{\rm BH})\sim 8.7$ and
$\dot{\rm m} \sim$ 5\% of the Eddington limit. The fact that the respective
Low-z sources are more variable can be explained by a smaller black hole mass
(log(M$_{\rm BH})\sim 8$) and a higher accretion rate ($\dot{\rm m} \sim$ 25\%
of the Eddington limit). At the same time, the higher accretion rate compensates
the difference in BH mass, and explains why objects with different BH mass 
have the same luminosity. 

The long dashed lines in the middle and bottom panels of Fig.~3 show the 
model  ``log($\sigma^2_{\rm NXS}$) vs log(L$_{\rm X}$)" lines for $\dot{\rm
m}_{\rm Edd,High-z}=0.4,\dot{\rm m}_{\rm Edd,vHigh-z}=0.5$, 
$\nu_{lf,\rm{High-z}}=1/20.2$ days$^{-1}$ and  $\nu_{lf,\rm{vHigh-z}}=1/12.7$
days$^{-1}$ (20.2 and 12.7 days being the average, rest-frame length of the
High-z and vHigh-z light curves, respectively). Clearly, these model curves 
describe well the mean variability amplitude vs X--ray luminosity data of these
objects. 

\subsection{The ``variability -- redshift" relation in AGN} According to the
model presented above, as the redshift increases, objects at a given luminosity
bin should correspond to AGN of smaller BH mass (hence the larger variability
amplitude) and higher accretion rate. Therefore, we would expect the
variability amplitude of AGN with the same luminosity to increase with
increasing redshift. In order to investigate this issue further, we considered
the ``variability -- redshift" relation at fixed luminosity bins. 

\begin{figure}
\centering
\includegraphics[height=8.5cm,width=8.5cm]{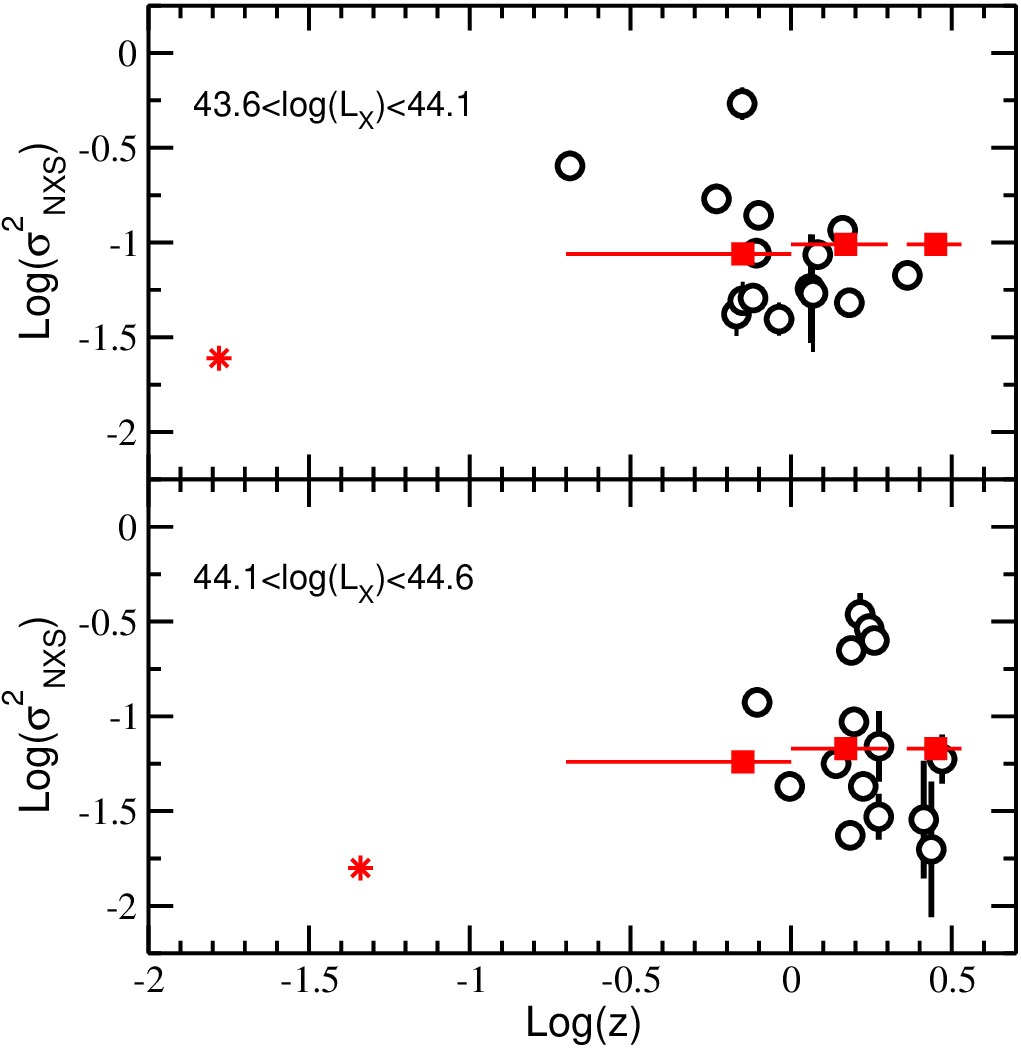}
\caption{The ``variability -- redshift" relation for the variable sources in
our sample with ``low" and ``high" luminosity (open circles in the  top and
bottom panels, respectively). Filled squares show the model predictions,
estimated as explained in Section 6.1, and stars indicate the average
variability amplitude and redshift of the ``lowLum" and ``highLum", 
nearby AGN.}
\end{figure}

Open circles in Fig.~4 shows the [log($\sigma^2_{\rm NXS}$),log(z)] data for
the variable objects in our sample with L$_X=10^{43.6}-10^{44.1}$ ergs/s (top
panel; the ``lowLum" bin) and L$_X=10^{44.1}-10^{44.6}$ ergs/s (bottom panel;
the ``highLum" bin). Star-like points in both panels show the average
variability amplitude and redshift for the nearby, ``lowLum" and ``highLum"
AGN. Contrary to the `variability -- redshift" anti-correlation we detect when
we consider all variable sources together (see top panel in Fig.~2), here one
can even claim a positive correlation between  $\sigma^2_{\rm NXS}$ and
redshift: the variability amplitude of the higher redshift, LH AGN is higher
than the amplitude of the nearby objects with the same luminosity. This is in
agreement with the discussion in the paragraph above. 

However, the situation is less clear when we consider the LH sources only 
(i.e. the AGN at redshift larger than $\sim 1$ or so). We find that there does
not exist a significant anti-correlation (or correlation) either in the ``low"
or ``highLum" bins ($\tau_{\rm lowLum}=-0.16$, $\tau_{\rm highLum}=-0.17$, with
P$_{\rm null,lowLum}=0.4$,  and P$_{\rm null,highLum}=0.37$, respectively). 

The average X--ray luminosity of the lowLum and highLum  sources is
$<\log(\rm{L}_{\rm X,lowLum})>=43.8$ and $<\log(\rm{L}_{\rm X,highLum})>=44.3$.
We can now use the $\dot{\rm m}_{\rm Edd}=0.25, 0.4$ and 0.5 model curves,
shown with the dashed lines in Fig.~3, to estimate the amplitude of objects
with L$_{\rm X}=10^{43.8}$ and L$_{\rm X}=10^{44.3}$  ergs/s in the Low-z,
High-z and vHigh-z redshift bins, respectively. Filled (grey) squares in Fig.~4
show the variability amplitude of these objects (as x-axis coordinates we have
used the average redshift of all sources in the Low-z, High-z and vHigh-z bins,
while the horizontal bars correspond to the size of the respective bins).
Clearly, the model ``variability - redshift" relations agree well with the data
in both luminosity bins.

The main reason that we do not observe a strong, positive correlation between
variability amplitude and redshift (when z$>$1) at fixed luminosity bins is
that, the rest frame light curve length, $T_{\rm rest frame}$, and hence
$\sigma^2_{\rm NXS}$ as well, decreases with increasing redshift. Therefore,
although the more distant AGN have a smaller  BH mass and higher accretion rate
than the local, same luminosity objects, their variability amplitude is not as
large as we would expect due to this  decrease in $T_{\rm rest frame}$.  This
``$T_{\rm rest frame}$ -- redshift" anti-correlation is strong enough to
compensate the increase in $\sigma^2_{\rm NXS}$ due to the higher $\dot{\rm
m}_{\rm Edd}$ and smaller M$_{\rm BH}$.

\section{Discussion and conclusions}

We have studied the long term X--ray variability properties of 66 AGN in the
Lockman Hole field which have optical spectroscopic identifications. We used
rest frame 2--10 keV band light curves, and we have taken particular care that
they all have a similar sampling pattern. For this reason, we extracted  data
from the last 10 \xmm\ observations of this region only, and we restricted our
sample to objects with N$_{\rm data}\ge 4$.

Using the $\chi^2$ test, we detected significant variations in 50 of them. The
low signal-to-noise ratio and the smaller number of points in the light curves
of the remaining 16 objects can explain the absence of significant variations
in them. We have used the well known ``normalized excess variance",
$\sigma^2_{\rm NXS}$, estimator  in order to quantify the variability amplitude
of these AGN. We have also used \rxte\ light curves of 12 nearby AGN, with an
average energy and length similar to the LH AGN, in order to measure their
normalized excess variance and compare their variability properties with the 
Lockman Hole AGN. Our results can be summarized as follows:

a) When we consider all the objects in our sample together, we detect a
significant anti-correlation between redshift, luminosity and variability
amplitude: the amplitude decreases with increasing redshift and luminosity.
This ``variability -- luminosity" anti-correlation is in agreement with (and
extends out to redshift of $\sim 3$) the results of Almaini \etal\ (2000), 
Manners \etal\ (2002) and Paolillo \etal\ (2004). 

b) We do not find a significant correlation between variability amplitude and
$\Gamma$. This is in agreement with the results of Mateos \etal\ (2007) but 
opposite to what has been observed in nearby AGN, where sources with steeper
spectra show larger amplitude variations (e.g. Green \etal\ 1993, Turner \etal\
1999, Grupe \etal\ 2001). Paolillo \etal\ (2004) also observed sources with
harder energy spectra to be less variable in their study of the X--ray
variability properties of the AGN in the {\it Chandra} Deep Field-South. In
order to investigate this issue further, we searched for a  ``variability --
$\Gamma$"  relation  in the Low-z, High-z and vHigh-z subsamples, but we did
not detect any. Perhaps, if real, such a  ``variability -- spectral shape"
correlation has such a large intrinsic scatter, in which case we would need a
larger sample to detect it. 

c) The  ``variability -- luminosity" relation of the Lockman Hole AGN has a
larger amplitude when compared to that of local AGN. This is in agreement with
the results of Almaini \etal\ (2000),  Manners \etal\ (2002) and Paolillo
\etal\ (2004), who also found that high-redshift sources behave differently, in
terms of their variability amplitude, than their nearby counterparts. 

d) The global ``variability amplitude--redshift/luminosity" anti-correlations
are less pronounced when we consider AGN in various luminosity and redshift
bins. The main trend we detect is that, at a given luminosity bin, the
variability amplitude increases with redshift until z$\sim$1, and then stays
roughly constant.

It is not possible to compare quantitatively our results with those presented
in the past. Quantitative comparisons can be performed when one uses light
curves of similar duration and rest frame energy band. Furthermore, the
variability amplitude must be measured in the same way, preferably with the use
of an estimator whose relation with the AGN fundamental physical parameters,
like the BH mass and accretion rate, can be established. Such an estimator is
the ``normalized excess variance", which we adopted in this work, since its
dependence on M$_{\rm BH}$, $\dot{{\rm m}}_{\rm Edd}$ {\it and} the light
curve's length can be easily determined (Section 6.1). 

In any case, we believe that parameters like the slope of the best fitting line
in the overall ``variability-- luminosity" or ``variability -- redshift" plots
(top and middle panels in Fig.~2) do not provide important physical insights.
The shape of the ``variability--luminosity/redshift" relations  is strongly
affected  by factors like the length of the light curves used and their energy
band. Such ``experimental" parameters can introduce or even ``destroy" any  
global ``variability-- luminosity" or ``variability -- redshift"
anti-correlations. For example, as we discussed in Section 4, although formally
significant, the ``variability-- luminosity" and ``variability -- redshift"
anti-correlations that we detect are mainly driven by a few points. A better
understanding of how the AGN variability properties change with look-back time
can be achieved by studying  the ``variability--luminosity" and/or the 
``variability--redshift" relations in small, fixed redshift and luminosity bins,
respectively. 

Our results are fully consistent with the assumption that the X--ray
variability mechanism and the accretion efficiency are the same in all AGN, at
all redshifts. In other words, our results strongly suggest that the X--ray
source operates in the same way in all AGN, at all redshifts. We find that,  as
the redshift increases, same luminosity AGN  have smaller BH mass and higher
accretion rates.  At the same time, as the redshift increases, the rest frame
light curve length  decreases accordingly. For this reason the variability
amplitude (for the same luminosity AGN) increases slightly up to z$\sim$1, and
then remains roughly constant with redshift. 

On the other hand, in the overall ``variability amplitude--redshift/luminosity"
plots (shown in Fig.~2) the high luminosity objects are mainly those with the
highest redshift as well. According to our results, their accretion rate should
be the highest among the sources in our sample. Consequently, they should also
show a large variability amplitude. However, they also have large BH masses,
and their {\it rest frame} light curve length is also small. These two effects
reduce significantly  the observed variability amplitude, and can explain the
global anti-correlation we observe between variability and
redshift/luminosity. 

\subsection{BH mass and accretion rate estimation} To the extent that all AGN
vary like the nearby ones (as our results suggest), we can provide BH mass and
accretion rate estimates for the objects in our sample. Obviously, given the
large scatter of the $\sigma^2_{\rm NXS}$ values, such a measurement will not
be very accurate if it were to be quoted individually for each object in our
sample. The variability method can only provide {\it average} estimates of
M$_{\rm BH}$ and $\dot{\rm m}_{\rm Edd}$. 

Even in this case, these estimates can only be considered as indicative at this
point.  For example, the largest number of sources (21) are in the ``Low-z"
bin. The $\dot{\rm m}_{\rm Edd}=0.4$ model curve in the middle panel of Fig.~2
appears to agree  well with the average [log($\sigma^2_{\rm nxs}), {\rm
log(L}_{\rm X})$] points in this bin, but this is hard to quantify.  With just
3 points to compare with, we cannot perform a proper $\chi^2$ fit to judge the
goodness of the model fit and provide confidence ranges for the best-fitting
model parameter values. We need significantly larger samples  of high-z AGN, in
order to improve the accuracy of our results.

Having these comments in mind, the model curves in Fig.~3 suggest that, on
average, the Low (z$\sim$0.7), High (z$\sim$1.5) and vHigh-z (z$\sim$2.9)
Lockman Hole AGN accrete at $\sim 25, 40$ and 50\% of the Eddington limit. Our
results are in agreement with those of McLure \& Dunlop (2004) who also observe
a slow evolution of the Eddington ratio  from $\dot{{\rm m}}_{\rm Edd}\sim
0.15$ at z$\sim 0.2$ to $\dot{{\rm m}}_{\rm Edd}\sim 0.5$ at z$\sim 2$, for a
large sample of quasars drawn from the SDSS catalogue. Although a few objects
in Fig.~3 show variability amplitudes well above the average (by a factor up to
10 in some cases),  these outliers are probably of stochastic nature (notice
for example that the scatter of points at a given luminosity in the middle
panel of Fig.~2 is of a similar magnitude). We conclude that, most probably,
the majority of the sources in our sample are accreting at rates below the
Eddington limit. This limit seems to be a relevant physical boundary to the AGN
accretion rate out to redshift $\sim 3$.

Regarding black hole masses, the model curves in Fig.~3 suggest a BH mass range
of $5\times 10^6 - 2\times 10^8$ M$_{\odot}$ in the case of the Low-z objects, 
$1.5\times 10^7 - 6.6\times 10^8$ M$_{\odot}$ for the High-z objects, and  
$5\times 10^7 - 1.3\times 10^9$ M$_{\odot}$ for the AGN in the vHigh-z bin.
Interestingly, even for these high-z and very luminous objects, the highest BH
mass estimate is less than $3\times 10^9$ M$_{\odot}$, i.e. the most massive BH
measured dynamically in the local Universe and the expected BH mass limit based
on the known properties of early-type galaxies and the locally observed
correlation between bulge and black hole mass (McLure \& Dunlop, 2004).

{\bf Appendix}

When dealing with variability, one of the main concerns  is the possible
contribution to the observed variations  of systematic effects due to
instrumental response and background. Subtle variations  of these instrumental
quantities may remain  uncorrected and can be difficult to spot.  In order to
verify that no systematic instrumental variations are introducing artifacts in
the light curves we use in this work, we divided the light curve of each
one of the V-objects (in the observer's frame) over its mean. Fig.~5 shows these
``normalized" light curves.  Most of the V-objects were mainly observed during
the first five and the last three of the 10 \xmm\ observations listed in
Table~2. This is due to the facts that a) we chose objects with N$_{\rm data}\ge
4$ and b) the shift between the pointing coordinates of the October 23 and 25
observations and the coordinates of the other eight observations is quite large.

We then estimated the mean of all the normalized count rates within each
observation (the ``mean normalized count rate" of the \xmm\ observations  are
listed in the second column of Table 3). If there are  any significant biases
introduced by instrumental effects in any of the \xmm\ observations, we would
expect the respective mean normalized count rate to be significantly different
than unity. It is only during the October 23observation that we observe the
mean to be higher than unity, but even in this case this difference is
significant at the $2.5\sigma$ level only. This result suggests that
instrumental effects do not affect significantly the observed light curves. 

\begin{figure}
\centering
\includegraphics[height=8.5cm,width=8.5cm]{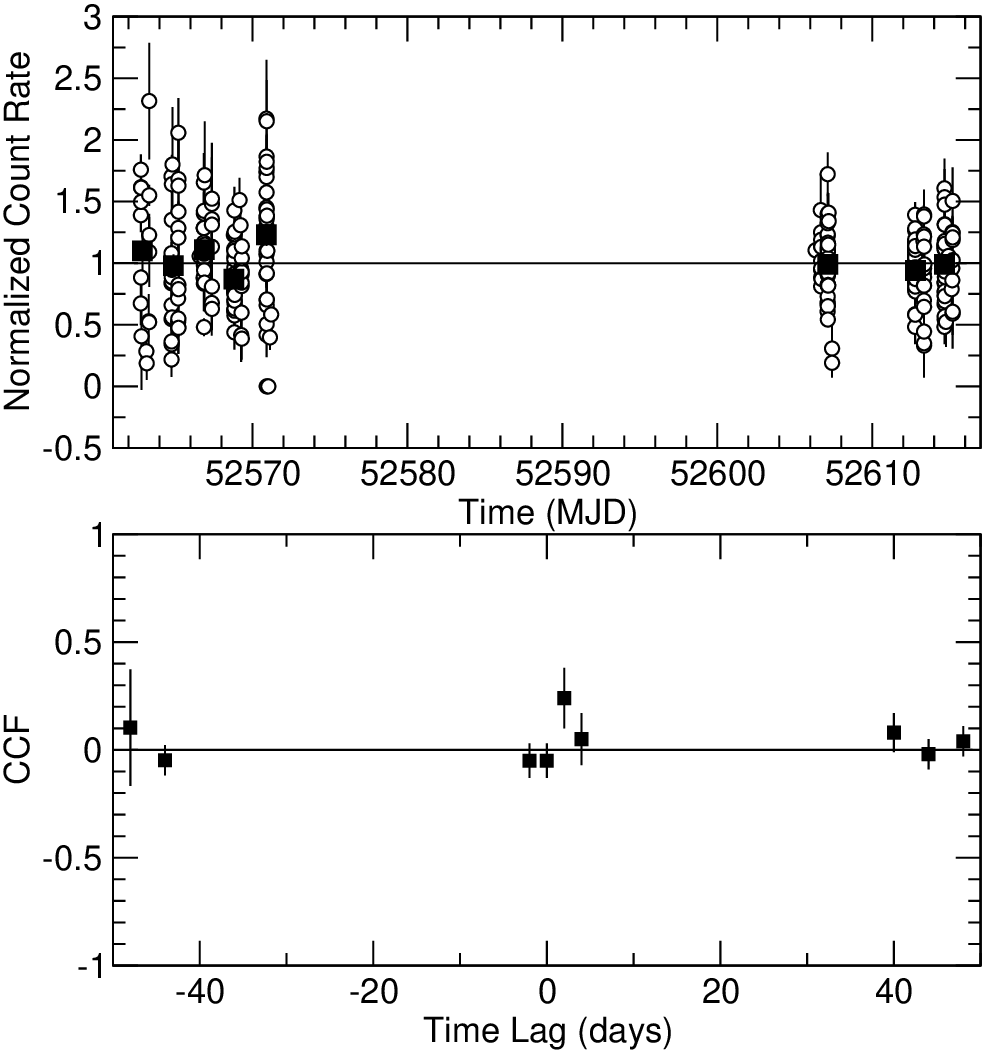}
\caption{Top panel: The normalized light curves of all the ``V" objects in our
sample (open circles). Time is in Modified Julian Days. For clarity reasons, we
have changed the x-coordinates randomly, by a small amount. Filled squares show
the mean of all the points within each observation (errors are also  plotted,
but are smaller than the symbol's size). Bottom panel: The average CCF between
source V148 and the light curves of all the other ``V" sources (in the observer's
frame).}
\end{figure}

This conclusion is further strengthened when we consider the cross correlation
between the light curves of the ``V" sources in our sample. In the bottom panel
of Fig.~5 we show the average cross-correlation function (CCF) between the V148
source and all the other V-sources (V148 is the brightest among the variable
sources with N$_{\rm data}=8$).  To construct this average CCF we computed first
the individual CCFs between the V148 and the other sources, using light curves
in the observer's frame, the DCF method of Edelson \& Krolik (1988) and a lags
bin of 2 days. We then computed the mean CCF at each lag where there were more
than 10 individual CCF estimates. There is no obvious significant structure in
the average CCF, which is rather flat, and consistent with zero at all lags.  We
reached the same result when we constructed the average CCFs using a few other
``reference sources". We therefore conclude that any unaccounted instrumental
effects should not contribute significantly to the variations we observe in the
V-sources light curves. 

\begin{table}
\begin{center}
\caption{The mean normalized count rate of all the ``V" objects in each \xmm\
observation, and the corresponding fractional root mean square variability
amplitude, F$_{\rm var}$.} 
\begin{tabular}{ccc}
\hline
\hline
Obs & Mean & F$_{\rm var}$ \\
\hline
2002--10--15 & 1.10($\pm$0.16)    &	 0.51($\pm$0.05)  \\
2002--10--17 & 0.98($\pm$0.09)    &	 0.44($\pm$0.05)  \\
2002--10--19 & 1.11($\pm$0.04)    &	 0.15($\pm$0.04)  \\
2002--10--21 & 0.87($\pm$0.05)    &	 0.25($\pm$0.03)  \\
2002--10--23 & 1.23($\pm$0.09)    &	 0.35($\pm$0.03)  \\
2002--11--27 & 0.99($\pm$0.04)    &  	 0.25($\pm$0.02)  \\
2002--12--04 & 0.94($\pm$0.04)    &      0.24($\pm$0.02)  \\
2002--12--06 & 0.99($\pm$0.04)    &	 0.24($\pm$0.02)  \\   
\hline  				  
\end{tabular}
\end{center}
\end{table}

The third column of Table~3 lists the ``root mean square (rms) variability
amplitude", F$_{\rm var}$, of all the normalized count rates within each
observation and its error (which accounts only for the effects of the flux
measurement errors). Both  the rms variability amplitude  and its error have been
estimated as in Vaughan \etal\ (2003). In effect, F$_{\rm var}$  measures the 
scatter of all the normalized count rates around their mean within each
observation. In theory, we should expect that F$_{\rm var}$ should remain roughly
constant: we can consider the flux measurements of the various V-sources in each
frame as an ensemble of different measurements of the same object. If the
variability process is stationary, then we would not expect the rms variability
amplitude to change significantly. However, this is meant to hold ``on average",
as F$_{\rm var}$ measurements from individual light curves of an object can easily
vary by a factor up to $\sim 3$ (see for example the bottom two panels in Fig.~6
of Vaughan \etal, 2003). 

The values listed in Table~3 show that F$_{\rm var}$ varies by a factor of $\sim
2-3$ from one observation to the other. It is difficult to understand how any
instrumental effects could introduce such a scatter in the F$_{var}$ values.
However,  even if this scatter in the F$_{\rm var}$ values is the result of the
red-noise character of the X-ray variability in AGN, we considered the possibility
it may affect the excess variance we measure for the faintest objects in our
sample. For that reason, we estimated the average F$_{\rm var}$ of the 15
brightest and faintest objects in the V-sample. More than 10 of them were detected
during the October 17 and 23 observations (which show large F$_{\rm var,total}$
values). We find that F$_{\rm var,bright,Oct. 17}=0.41\pm 0.05$ and  F$_{\rm
var,bright,Oct. 23}=0.38\pm 0.05$, while  F$_{\rm var,faint,Oct. 17}=0.35\pm 0.11$
and  F$_{\rm var,faint,Oct. 23}=0.28\pm 0.10$. Given the fact that the fainter
objects show smaller scatter around their mean, we are confident that instrumental
effects do not introduce significant ``artificial" variations in the light curves
we use in this work.

\end{document}